\documentclass[twoside,fleqn]{article}
\usepackage{espcrc2}
\usepackage{graphics}
\usepackage{epsfig}

\def\CO{{\cal O}}

\def\Str{\textrm{Str}}

\def\etal{{\it et al.}}

\def\spose#1{\hbox to 0pt{#1\hss}}
\def\ltapprox{\mathrel{\spose{\lower 3pt\hbox{$\mathchar"218$}}
 \raise 2.0pt\hbox{$\mathchar"13C$}}}
\def\gtapprox{\mathrel{\spose{\lower 3pt\hbox{$\mathchar"218$}}
 \raise 2.0pt\hbox{$\mathchar"13E$}}}
\def\inapprox{\mathrel{\spose{\lower 3pt\hbox{$\mathchar"218$}}
	\raise 2.0pt\hbox{$\mathchar"232$}}}

\begin{document}

\title{
       \begin{flushright}\normalsize
	    \vskip -1 cm
            UW/PT 04-14
       \end{flushright}
	\vskip -0.4 cm
Next-to-Leading-Order Staggered Chiral Perturbation Theory
\thanks{Presented by R.~Van~de~Water.
Research supported in part by US-DOE contract DE-FG02-96ER40956.}
}

\author{Stephen R. Sharpe$\rm ^{a}$ and Ruth S. Van de Water\address{Physics Department, Box 351560,
University of Washington, Seattle, WA 98195-1560, USA}}

\begin{abstract}

\vspace{-.10in}We have extended staggered chiral perturbation theory (S$\chi$PT) to $\CO(a^2 p^2)$, $\CO(a^4)$, and  $\CO(a^2 m)$, the orders necessary for a full next-to-leading order (NLO) calculation of pseudo-Goldstone boson (PGB) masses and decay constants including taste-symmetry violations.  We present predictions relating $SO(4)$ taste-breaking splittings in masses, pseudoscalar decay constants, and dispersion relations.  These can be used to test the $\sqrt[4]{\textrm{Det}}$ trick.

\vspace{-.20in}\end{abstract}

\maketitle

\section{Introduction}

Lattice simulations with staggered fermions\cite{Susskind} are fast compared to those with other standard fermion discretizations\cite{BerlinPanel,Daviesetal}.  They therefore allow QCD calculations on a reasonable time scale with light dynamical quarks. In addition to this computational advantage, staggered fermions also possess a remnant $U(1)$ chiral symmetry, even at nonzero lattice spacing.  Their primary disadvantage is that each continuum staggered fermion flavor comes in four degenerate tastes, and the continuum $SU(4)$ symmetry associated with each flavor is almost completely broken by the lattice, leaving only a discrete subgroup.  While taste-breaking effects only enter at $\CO(a^2)$, they turn out to be important numerically at current lattice spacings\cite{MILCspectrum}.  

In order to extract physical quantities from lattice simulations, one must perform both continuum and chiral extrapolations using functional forms determined in chiral perturbation theory\cite{MILCf}.  Thus, forms applied to staggered simulations must account for taste violations.  Near the continuum limit, where there is a controlled expansion in powers of the lattice spacing, S$\chi$PT systematically describes taste violations in the PGB sector.  Reference~\cite{LS} determined the $\CO(a^2)$ staggered potential for a single flavor and showed that even at tree-level it correctly predicts the mass degeneracy pattern observed among the different tastes of PGBs\cite{JLQCD1,JLQCD2,OT}.  References~\cite{BA1} and~\cite{BA2} extended the potential to multiple staggered flavors and calculated the 1-loop mass and decay constant for the taste Goldstone ($\xi_5$) meson.  However, it is well known from continuum $\chi$PT that one must work to at least NLO for adequate chiral extrapolation of lattice data\cite{MITpanel}.  Thus, the utility of S$\chi$PT is limited without higher-order operators.  We have therefore determined all linearly-independent $\CO(a^2 p^2)$, $\CO(a^4)$, and $\CO(a^2m)$ operators in the staggered chiral Lagrangian for a general multiple-flavor, partially quenched theory, including source terms for left- and right-handed currents and scalar and pseudoscalar densities.\footnote{The full list of NLO taste-breaking operators is given elsewhere\cite{Us}.}  The full NLO staggered chiral Lagrangian consists of these operators, which incorporate discretization effects, plus the standard continuum $\CO(p^4)$, $\CO(p^2 m)$, and $\CO(m^2)$ operators. 

\vspace*{-1.4mm}\section{NLO Staggered Chiral Lagrangian}

Determination of the staggered chiral Lagrangian requires two steps.  First one constructs the quark level effective action, including explicit nonzero $a$ effects.  Following the method of Symanzik\cite{Symanzik}, $S_{eff}$ consists of all local continuum lattice operators of dimension $\leq$ 6 invariant under the lattice symmetries.  In practice it is easiest to first determine all allowed \emph{lattice} operators of dimension $\leq$ 6, as in Refs.~\cite{Luo,LS}, and then map these onto continuum quark-level operators\cite{LS}.  Next one maps the continuum quark-level operators onto chiral operators describing the PGB sector.  To be most general, we consider the partially quenched (PQ) theory -- unquenched QCD can be recovered by taking an appropriate limit.   S$\chi$PT uses the following power counting scheme: $p^2/\Lambda_{QCD}^2 \approx m/\Lambda_{QCD} \approx a^2 \Lambda_{QCD}^2$, which applies to current staggered lattice simulations.  Thus the LO staggered potential is of $\CO(a^2)$\cite{LS,BA1}, while the NLO taste-violating operators in the staggered chiral Lagrangian are of $\CO(a^2 p^2)$, $\CO(a^4)$, and $\CO(a^2 m)$\cite{Us}.  

The NLO chiral operators can be divided into two types according to how badly they break the continuum $SU(4)$ taste symmetry.  Those of type A are rotationally invariant and preserve an $SO(4)$ subgroup of the continuum $SU(4)$ taste symmetry.  In contrast, type B operators are only invariant under certain combined spin and taste rotations, and \emph{maximally} break the continuum symmetry down to the lattice symmetry group.  Because all operators in the LO chiral Lagrangian are of type A, the tree-level PGB masses are split into irreps of $SO(4)$-taste:  $\xi_I$, $\xi_5$, $\xi_\mu$, $\xi_{\mu 5}$, and $\xi_{\mu \nu}$.  However, analytic NLO contributions due solely to the B-type operators can cause further mass and decay constant splittings among the tastes.

\vspace*{-1.2mm}\section{NLO Relations for PGB Properties}

Because there are 234 new NLO taste-breaking operators, determining all of their coefficients does not seem possible.  Nevertheless, there is one set of quantities to which only a few contribute and for which predictions are possible -- those which are nonzero only because $SO(4)$-taste symmetry is broken at NLO.  

We first consider $SO(4)$-taste and rotational symmetry violation in the pion dispersion relations.  Only eight $\CO(a^2 p^2)$ operators contribute\footnote{Operators of $\CO(a^2 m)$ only occur in type A, while type B $\CO(a^4)$ operators contract into $SO(4)$-invariant ones for two-PGB processes.}: 
\begin{eqnarray}
	&& \!\!\!\!\!\! a^2 \; \sum_\mu \; \sum_{\nu \ne \mu} \; \big\{ C_2 \Str(\partial_\mu
\Sigma^\dagger \xi_{\mu\nu} \partial_\mu \Sigma \xi_{\nu\mu})  \nonumber \\
	&& \!\!\!\!\!\!\! + \:C_7\: \Str( \Sigma \partial_\mu \Sigma^\dagger \xi_{\mu\nu}) \Str( \Sigma^\dagger \partial_\mu \Sigma \xi_{\nu\mu})  \nonumber \\
	&& \!\!\!\!\!\!\! + \;C_{10} \left[\Str( \Sigma\partial_\mu \Sigma^\dagger \xi_{\mu\nu} \Sigma \partial_\mu \Sigma^\dagger \xi_{\nu\mu}) + p.c. \right] \nonumber \\
	&& \!\!\!\!\!\!\! + \:C_{13}  \left[ \Str( \Sigma \partial_\mu \Sigma^\dagger \xi_{\mu\nu}) \Str( \Sigma \partial_\mu \Sigma^\dagger \xi_{\nu\mu}) + p.c. \right] \big\} \nonumber \\ 
	&& \!\!\!\!\!\!\!+ \; a^2 \; \sum_\mu \; \big\{ C_{44V}\left[ \Str(\partial_\mu \Sigma^\dagger
\xi_\mu \partial_\mu \Sigma^\dagger \xi_\mu) + p.c. \right] \nonumber \\
	&& \!\!\!\!\!\!\! + \:C_{44A}\left[ \Str(\partial_\mu \Sigma^\dagger \xi_{\mu 5} \partial_\mu \Sigma^\dagger \xi_{5 \mu}) + p.c. \right]   \nonumber \\
	&& \!\!\!\!\!\!\! + \;C_{47V} \left[\Str(\partial_\mu \Sigma^\dagger\xi_\mu)^2 + p.c.\right] \nonumber \\
	&& \!\!\!\!\!\!\! + \:C_{47A} \left[\Str(\partial_\mu \Sigma^\dagger\xi_{\mu 5})
	\Str(\partial_\mu\Sigma^\dagger \xi_{5 \mu}) + p.c.\right] \big\}. 
\end{eqnarray}
One can see that these operators break $SO(4)$-taste and Euclidean rotational symmetry from the fact that they have more than two repeated indices.  This property is common to all B-type operators in the staggered chiral Lagrangian.  Note that four of these operators contain two supertraces;  they produce hairpin correlators which only contribute to properties of flavor-singlet PGBs.  Taste $\xi_{\mu \nu}$ and $\xi_5$ hairpins were not present in the LO Lagrangian, but do occur at NLO.

As expected, we observe rotational symmetry violation in the pion dispersion relations\footnote{Dispersion relations for tastes I and 5 are Euclidean rotation invariant at this order.}: 
\begin{eqnarray}
	E^2_{k} &\!\!\!\!  =  &  \!\!\!\!(p_i^2 \!+ p_j^2) \!+ p_k^2 (1 \!+ \delta_k \!- \delta_4) \!+ m_\mu^2 (1 \!+ \delta_k), \nonumber \\
	E^2_{4} &\!\!\!\! = &  \!\!\!\!\vec{p}{\;}^2 (1 \!+ \delta_4 \!- \delta_k) +\, m_\mu^2 (1 \!+ \delta_4)\,, \nonumber \\
	E^2_{{k5}} &\!\!\!\!  = & \!\!\!\!(p_i^2 \!+ p_j^2) \!+ p_k^2 (1 \!+ \delta_{k5} \!- \delta_{45}) \!+ m_{\mu5}^2 (1 \!+ \delta_{k5}), \nonumber \\
	E^2_{{45}} &\!\!\!\!  =  &  \!\!\!\!\vec{p}{\;}^2 (1 \!+ \delta_{45} \!- \delta_{k5}) \!+ m_{\mu5}^2 (1 \!+ \delta_{45}) \,, \nonumber \\
	E^2_{{lm}} &\!\!\!\!  = & \!\!\!\!(p_l^2 \!+ p_m^2) (1 \!+ \delta_{lm} \!- \delta_{k4}) \nonumber \\
		&& +\; p_k^2 + m_{\mu \nu}^2 (1 \!+ \delta_{lm}), \nonumber \\
	E^2_{{k4}} &\!\!\!\!  = & \!\!\!\!(p_l^2 \!+ p_m^2)(1 \!+ \delta_{k4} \!- \delta_{lm}) \nonumber \\
		&& +\; p_k^2 + m_{\mu \nu}^2 (1 \!+ \delta_{k4}) \,.
\end{eqnarray}
Here $E_F$ is the energy of a taste $F$ pion extracted from the exponential fall-off of the 2-point function along the Euclidean time (4) direction.  The masses, $m_F$, are the full NLO masses \emph{excluding} the contributions from the type B operators enumerated above, so they fall into irreps of $SO(4)$.  The corrections, $\delta_F = \delta m_F^2 / m_F^2$ arise from the additional $SO(4)$-breaking operators, and further split the tastes into irreps of the \emph{timeslice group}, the rest frame symmetry group of the staggered action\cite{Golterman}\footnote{Explicit expressions for the $\delta$s are not needed for the 
predictions and are given elsewhere.\cite{Us}.}.  As this is the maximum possible splitting among tastes on a lattice with zero spatial momentum, these degeneracy classes hold to all orders in S$\chi$PT.  A testable prediction can be made in terms of mass and energy splittings among the $SO(4)$ irreps $\xi_\mu$, $\xi_{\mu 5}$, and $\xi_{\mu \nu}$:
\begin{eqnarray}
	\frac{E_k^2 - E_4^2}{m_k^2 - m_4^2} &\!=& \!1 + \frac{p_i^2 + p_j^2 + 2 p_k^2}{(m_k^2+m_4^2)/2}, \nonumber \\
	\frac{E_{k5}^2 - E_{45}^2}{m_{k5}^2 - m_{45}^2} &\!=& \!1 + \frac{p_i^2 + p_j^2 + 2 p_k^2}{(m_{k5}^2+m_{45}^2)/2}, \nonumber \\
	\frac{E_{lm}^2 - E_{k4}^2}{m_{lm}^2 - m_{k4}^2} &\!=& \!1 + 2 \frac{p_l^2 + p_m^2}{(m_{lm}^2+m_{k4}^2)/2}.
\end{eqnarray}
These are now the full NLO masses which are directly measurable on the lattice.

The only Euclidean rotational and $SO(4)$ taste symmetry-violating contributions to both PGB masses and pseudoscalar decay constants are from the same $\CO(a^2 p^2)$ operators, through wavefunction renormalization.  Thus the splittings within $SO(4)$ irreps in these two quantities are related:  
\begin{eqnarray}
	\Big(\frac{f^P_k - f^P_4}{f^P_k + f^P_4}\Big) & = & \frac{1}{2} \Big(\frac{m_k^2 - m_4^2}{m_k^2 + m_4^2}\Big) \,, \nonumber \\
	\Big(\frac{f^P_{k5} - f^P_{45}}{f^P_{k5} + f^P_{45}}\Big) & = & \frac12 \Big(\frac{m_{k5}^2 - m_{45}^2}{m_{k5}^2 + m_{45}^2}\Big) \,, \nonumber \\
	\Big(\frac{f^P_{lm} - f^P_{k4}}{f^P_{lm} + f^P_{k4}}\Big) & = & \frac12 \Big(\frac{m_{lm}^2 - m_{k4}^2}{m_{lm}^2 + m_{k4}^2}\Big)\,. 
\end{eqnarray}
This is the simplest prediction of NLO S$\chi$PT that one can test on the lattice.  It is essential for these expressions that the $Z$-factors are $SO(4)$-invariant, and therefore identical for both tastes in the expression.  Thus they can be tested using \emph{bare lattice operators}, thereby avoiding $\CO(a^2)$ ambiguities in matching lattice and continuum operators that could destroy the relationship.  Because, in the $\delta$s, there are three independent coefficients, and there are three splittings, there are no predictions relating mass splittings among different $SO(4)$ irreps.

We note that, in the talk at Lattice 2004, we presented a different prediction involving the \emph{axial} decay constant:
\begin{equation}
	\Big(\frac{f^A_k - f^A_4}{f^A_k + f^A_4}\Big) = -\Big(\frac{m_k^2 - m_4^2}{m_k^2 + m_4^2}\Big)\,, 
\end{equation}
and so forth.  We now realize that this is incorrect for two reasons.  First, it does not properly take into account wavefunction renormalization when calculating $f^A$.  Moreover, even if one correctly calculates the contributions from the listed $\CO(a^2 p^2)$ operators, this relationship is broken by additional operators containing sources for left- and right-handed currents.  These operators, in which the covariant derivative acts on a taste spurion, rather than a $\Sigma$-field, affect the axial current, but not the masses or pseudoscalar density.

\vspace*{-1.4mm}\section{Conclusion}

We have used NLO S$\chi$PT to predict {\emph{quantitative relationships}} between PGB masses, decay constants, and dispersion relations.  We emphasize that, although the underlying lattice symmetries tell us that the approximate degeneracies among the $SO(4)$ irreps must be split into the true lattice irreps, they do not predict any particular relationships among the $SO(4)$ splittings in various PGB properties.  These come strictly from S$\chi$PT. These relationships therefore provide a {\emph{simple test} of whether the effective field theory applies \emph{at all} to the lattice data, which is based on the assumption that the underlying lattice ``theory'' using the $\sqrt[4]{\mbox{Det}}$ trick is not sick.  Testing them will provide empirical evidence either for or against the validity of this trick.

\end{document}